\documentclass[12pt]{article}
\usepackage{amsfonts}
\usepackage{graphicx}
\usepackage{amssymb}
\usepackage{amsmath}
\begin{document}
\renewcommand{\thefootnote}{\fnsymbol{footnote}}
\begin{titlepage}

\vspace{10mm}
\begin{center}
{\Large\bf Spontaneous breaking of permutation symmetry in pseudo-Hermitian quantum mechanics}
\vspace{16mm}

{\large Jun-Qing Li${}^{1}$ and Yan-Gang Miao${}^{1,2,3,}$\footnote{Corresponding author. E-mail address: miaoyg@nankai.edu.cn}

\vspace{6mm}
${}^{1}${\normalsize \em School of Physics, Nankai University, Tianjin 300071, China}

\vspace{3mm}
${}^{2}${\normalsize \em Kavli Institute for Theoretical Physics China, CAS, Beijing 100190, China}

\vspace{3mm}
${}^{3}${\normalsize \em Bethe Center for Theoretical Physics and Institute of Physics, University of Bonn, \\
Nussallee 12, D-53115 Bonn, Germany}}

\end{center}

\vspace{10mm}
\centerline{{\bf{Abstract}}}
\vspace{6mm}
\noindent
By adding an imaginary interacting term proportional to $ip_1p_2$ to the Hamiltonian of a free anisotropic
planar oscillator, we construct a new model which
is described by the ${PT}$-pseudo-Hermitian Hamiltonian with the permutation symmetry of two dimensions.
We prove that our model is equivalent to the Pais-Uhlenbeck oscillator 
and thus establish a relationship between our $PT$-pseudo-Hermitian system and the fourth-order derivative oscillator model.
We also point out the spontaneous breaking of permutation symmetry which plays a crucial role in giving a real spectrum free of interchange of positive
and negative energy levels in our model.
Moreover, we find that the permutation symmetry of two
dimensions in our Hamiltonian
corresponds to the identity (not in magnitude but in attribute) of two different frequencies in the Pais-Uhlenbeck oscillator, and reveal that the
unequal-frequency condition imposed as a prerequisite upon the Pais-Uhlenbeck oscillator can reasonably be explained as the
spontaneous breaking of this identity.

\vskip 20pt
\noindent
{\bf PACS Number(s)}: 03.65.-w; 03.65.Fd; 03.65.Ta


\end{titlepage}
\newpage
\renewcommand{\thefootnote}{\arabic{footnote}}
\setcounter{footnote}{0}
\setcounter{page}{2}

\section{Introduction}

A non-Hermitian Hamiltonian had not attracted much attention because
it was regarded in general from the conventional point of view as an unobservable without real eigenvalues.
In the early 1940's, a non-Hermitian Hamiltonian and its associated indefinite metric in the Hilbert space were
introduced for the first time by Dirac~\cite{Dirac} and Pauli~\cite{Pauli} to deal with some
divergence problems related to some of the most fundamental physical concepts,
such as the unitarity of time evolution (conservation of probability). After thirty years
this idea was
applied to the quantum electrodynamics by Lee and Wick~\cite{LW} to keep the unitarity of $S$-matrix.
Later, some other work~\cite{BFMC} in different areas of research revealed
that a non-Hermitian Hamiltonian could have real eigenvalues under specific conditions.
It is worthy mentioning that a special class of non-Hermitian Hamiltonians, i.e. the $PT$-symmetric
Hamiltonian~\cite{BBR} has led to a long-lasting interest to the research of non-Hermitian Hamiltonians till now.
In ref.~\cite{BBR}, the $PT$-symmetric
Hamiltonian, $H=p^2+x^2(ix)^\epsilon$, is constructed and the relationship
between the ${PT}$-symmetry and real spectra is clarified in detail,
where $\epsilon$ is a real parameter, and ${P}$ and  ${T}$ stand for the space reflection and time reversal transformations, respectively.

Since the $PT$-symmetric Hamiltonian~\cite{BBR} was proposed,
a great progress has been made on the quantum mechanics
related to non-Hermitian Hamiltonians, i.e. the so-called non-Hermitian quantum mechanics, in contrast to the conventional (Hermitian) quantum mechanics.
See, for instance, some  review articles~\cite{CBM}.
The non-Hermitian quantum mechanics now mostly represents the ${PT}$-symmetric
quantum mechanics~\cite{BBR} and $\eta$-pseudo-Hermitian quantum mechanics~\cite{Pauli,MP}.
We should mention that the idea of non-Hermitian Hamiltonians has spread to the
quantum field theory~\cite{BBJ}, the supersymmetric quantum mechanics and quantum field
theory~\cite{DDT}, the noncommutative field theory~\cite{YGMCB}, the biological physics~\cite{EME}, and the quantum information~\cite{JS}, etc.
Most recently, the $PT$-symmetric  quantum mechanics
has been extended~\cite{JSM} to fermions with the odd time reversal.

If a non-Hermitian Hamiltonian $H$ satisfies the relation:\footnote{The ${PT}$-symmetry is written usually as
$H=H^{\rm PT} := (PT)H(PT)^{-1}$ or simply as $H=H^{\rm PT} := (PT)H(PT)$. We choose their equivalent
form, eq.~(\ref{PT}), in order to unify it to the form of $\eta$-pseudo-Hermitian symmetry, see eq.~(\ref{pseudoH}).}
\begin{equation}
H=H^{\rm PT} := (PT)^{-1}H(PT),\label{PT}
\end{equation}
that is, $H$ is ${PT}$-symmetric, its eigenvalues correspond to
complex conjugate pairs. Furthermore, if the ${PT}$-symmetry of $H$ is unbroken,
this Hamiltonian definitely
has a real spectrum. Besides eigenvalues some other fundamental problems have also been investigated, such as the positive definite inner
product. In a ${PT}$-symmetric quantum system, the inner product would not be positive definite
if it were defined just in terms of the $PT$ operation. Instead, a previously hidden  $C$ symmetry~\cite{BBJ2} should be combined.
The inner product modified by the ${CPT}$ operation is thus positive definite, and correspondingly the unitarity of  time evolution is
ensured. Therefore, the $PT$-symmetric
quantum mechanics possesses all the
desirable qualities normally existed in the conventional (Hermitian) quantum mechanics.

The $\eta$-pseudo-Hermitian quantum mechanics
was in fact proposed first by Pauli~\cite{Pauli} in the early 1940's.
If it satisfies a class of pseudo-Hermitian symmetries,\footnote{It was simply called self-adjoint by Pauli~\cite{Pauli}.
This self-adjoint condition ensures the conservation of probability in the Hilbert space
for any linear Hermitian operator $\eta$, where the probability is defined with respect to the indefinite metric  $\eta$, see ref.~\cite{Pauli}.}
\begin{equation}
H=H^{\ddag} := \eta^{-1} H^\dagger\eta,\label{pseudoH}
\end{equation}
where the dagger means the Hermitian adjoint as usual and $\eta$ is a linear Hermitian operator
or an anti-linear anti-Hermitian operator,\footnote{The anti-linear operator $\eta$ satisfies:
$\eta(a|\psi_1\rangle+b|\psi_2\rangle)=\overline{a}\eta|\psi_1\rangle+\overline{b}\eta|\psi_2\rangle$, where $a$ and $b$ are complex parameters and
the overline stands for complex conjugate. According to the definition given in ref.~\cite{Wein}, the anti-linear operator is said to be anti-Hermitian
if it satisfies:
$\langle\psi_1|\eta|\psi_2\rangle=\langle\psi_2|\eta|\psi_1\rangle$. For instance, the time reversal operator $T$ is
such an anti-linear anti-Hermitian operator. As the parity operator $P$ is linear Hermitian, the combined operator $PT$ is
thus anti-linear anti-Hermitian like $T$.}
$H$ describes an $\eta$-pseudo-Hermitian quantum
system~\cite{Pauli,MP}.
As stated in ref.~\cite{MP},
among the operators $\eta$ satisfying eq.~(\ref{pseudoH}) there exists under certain conditions
a class of linear Hermitian operators
with which the positive definite inner product can be introduced:
$\langle\psi|\phi\rangle_{\eta}:=\langle\psi|{\eta}|\phi\rangle$, and furthermore such a positive operator can actually be constructed.
The Hamiltonian has real average values with respect to such a positive definite inner product. Therefore,
a consistent quantum theory that complies with the requirement of a Hermitian quantum theory, such as the real eigenvalues,
the positive definite inner product and the unitary time evolution, can be established. However,
for an anti-linear anti-Hermitian $\eta$, the $\eta$-pseudo-Hermiticity
gives weaker restrictions to the Hamiltonian system.
That is, some properties (constraints) that exist in the case of a linear Hermitian $\eta$ disappear
in the case of an anti-linear anti-Hermitian $\eta$. For the details, see the next section and Appendix.

Recently the Pais-Uhlenbeck oscillator~\cite{PU} was studied as a $PT$-symmetric system
by means of the imaginary-scaling scheme~\cite{BM1,BM2} in which this
fourth-order derivative model has been shown to be stable and unitary. This leads us to our motivation in the present paper to give
in an alternative way an $\eta$-pseudo-Hermitian Hamiltonian that can describe the Pais-Uhlenbeck oscillator.
We indeed find a $PT$-pseudo-Hermitian system that includes the Pais-Uhlenbeck oscillator as a special case. More importantly,
we find that our $PT$-pseudo-Hermitian model has the permutation symmetry and that its spontaneous breaking plays a crucial role in giving
a real spectrum free of interchange of positive
and negative energy levels in our model. Furthermore, considering the equivalence of our model and the Pais-Uhlenbeck oscillator,
we give a suitable explanation for the unequal-frequency condition taken in ref.~\cite{BM1,BM2}, i.e., the source of such a condition
is the spontaneous breaking of the identity (not in magnitude but in attribute) of two frequencies happened
in the unequal-frequency Pais-Uhlenbeck oscillator.

In the next section we shall construct our $PT$-pseudo-Hermitian model, see eq.~(\ref{hamilton}).
Because the operator $PT$ is anti-linear anti-Hermitian (see footnote 3),
we also briefly discuss some properties that an $\eta$-pseudo-Hermitian quantum system
related to an anti-linear anti-Hermitian $\eta$ has. Then in section 3 we prove the equivalence of our model and the Pais-Uhlenbeck oscillator.
Next, we analyze the permutation symmetry of two dimensions and its spontaneous breaking in our model,
and point out the identity of two frequencies and its spontaneous breaking in the
Pais-Uhlenbeck oscillator in section 4.
Finally we make a brief conclusion in section 5.

\section{Our model and some properties of pseudo-Hermitian systems}

We add a non-Hermitian interacting term proportional to $ip_1p_2$ to the Hamiltonian of a free anisotropic planar
oscillator, and then give a new Hamiltonian:
\begin{equation}
H=\frac{1}{2}\left(p^2_1+p^2_2\right)+\frac{1}{2}\left(a^2_1 x^2_1+a^2_2
x^2_2\right)+i\frac{a_3}{2a_1 a_2}p_1p_2,\label{hamilton}
\end{equation}
where the non-vanishing constants $a_1$, $a_2$ and $a_3$ are real, and $a_1\neq a_2$; $(x_j, p_j)$, $j=1,2$,
are two pairs of canonical coordinates and their
conjugate momenta, they are all Hermitian and satisfy the standard Heisenberg commutation relations:
\begin{equation}
[x_j,p_k]=i{\delta}_{jk}, \qquad [x_j,x_k]=0=[p_j,p_k], \qquad
j,k=1,2,\label{canonicalRe}
\end{equation}
where $\hbar$ is set to be unity through out this paper.
When the conventional definitions of $P$ and $T$ are applied,\footnote{In general, there are many possible ways to define the parity $P$
as long as $P^2=1$ and $P$ is Hermitian. However, it is better to apply the same definition in a multi-dimensional coordinate system and
the conventional definition in a one-dimensional coordinate system in order to avoid inconsistency. Here we give two examples where the
inconsistency occurs. In a
two-dimensional coordinate system, if the conventional and unconventional definitions of $P$ are adopted for $x$ and $y$, respectively,
i.e., $x \rightarrow -x$;
$y \rightarrow y$, the quadratic models studied in section 3 of ref.~\cite{BJ} are no longer $PT$-symmetric. Another example for a
one-dimensional case, if the unconventional definition is utilized, $x \rightarrow x$, the well-known $PT$ symmetric Hamiltonian,
$H=p^2+x^2+ix$, is no longer $PT$-symmetric, either.}
that is,
\begin{eqnarray}
P: \qquad x_j \rightarrow -x_j, \qquad p_j \rightarrow -p_j, \qquad i \rightarrow +i;\nonumber \\
T: \qquad x_j \rightarrow +x_j, \qquad p_j \rightarrow -p_j, \qquad i \rightarrow -i,\label{PTdef}
\end{eqnarray}
it is obvious that the Hamiltonian eq.~(\ref{hamilton}) is neither Hermitian, $H \neq H^\dagger$, nor $PT$-symmetric, $H \neq (PT)^{-1}H(PT)$.
Instead, it possesses the pseudo-Hermitian symmetry defined by eq.~(\ref{pseudoH}),
where $\eta=PT$ is an anti-linear anti-Hermitian operator due to the time reversal $T$.
That is to say, $H$ has the ${PT}$-pseudo-Hermiticity or ${PT}$-pseudo-Hermitian symmetry, or in other words,
$H$ is ${PT}$-pseudo-Hermitian self-adjoint,\footnote{If $\eta=P_1, P_2$, or $T$, the Hamiltonian
(eq.~(\ref{hamilton})) also has the corresponding $\eta$-pseudo-Hermitian symmetry, where $P_1$ means the reflection of the first dimension
of coordinates and momenta while $P_2$ the reflection of the second dimension. Here we present the Hamiltonian
the maximal pseudo-Hermitian symmetry combined by $P_1$, $P_2$ and $T$.}
\begin{equation}
H=H^{\ddag} := (PT)^{-1} H^\dagger (PT).\label{PTpseudo}
\end{equation}

Here we compare the coupled quadratic Hamiltonians~\cite{BJ} with ours. In ref.~\cite{BJ} the interaction of a $PT$-symmetric Hamiltonian
and a Hermitian Hamiltonian is investigated, where the given
quadratic interaction is in fact a cross term of two spatial dimensions. Although the interacting term is quadratic
in both the Hamiltonians of ref.~\cite{BJ} and our Hamiltonian eq.~(\ref{hamilton}),
there are three differences between  them: (a) With the conventional definitions of $P$ and $T$, see eq.~(\ref{PTdef}), the former is $PT$-symmetric,
while the latter is  $PT$-pseudo-Hermitian; (b) The interacting term is real and quadratic in spatial variables  in ref.~\cite{BJ},
while it is imaginary and quadratic in momental variables in eq.~(\ref{hamilton});
(c) The interacting term is Hermitian in the Hamiltonians of ref.~\cite{BJ}, while it is $PT$-pseudo-Hermitian
in eq.~(\ref{hamilton}).
Moreover, in ref.~\cite{BJ} the relationship between a real spectrum and a coupling constant is discussed, while in the present paper
the equivalence of our model and the Pais-Uhlenbeck oscillator is revealed, and the significance of
the spontaneous breaking of permutation symmetry 
is pointed out for giving a real spectrum free of interchange of positive
and negative energy levels.

Following Pauli~\cite{Pauli} for a linear Hermitian operator $\eta$ as the indefinite metric,
we can discuss the $\eta$-pseudo-Hermitian quantum system
related to an anti-linear anti-Hermitian $\eta$. In particular, we have interest when $\eta=PT$.
As briefly mentioned under eq.~(\ref{pseudoH}), we can see more clearly in the following a great difference that
some fundamental properties for a consistent quantum system
no longer exist in the case of an anti-linear anti-Hermitian $\eta$.
For the details, see the Appendix. Here we just list the main results.
\begin{enumerate}
\item
A linear Hermitian $\eta$ leads to a real probability, i.e. the real bilinear form of wavefunctions,
$\overline{\langle\psi|{\eta}|\psi\rangle}=\langle\psi|{\eta}|\psi\rangle$,
while an anti-linear anti-Hermitian $\eta$ does not in general. Note that our model associated with
the anti-linear anti-Hermitian $PT$ is a special case. See section 4 for the details.

\item
A linear Hermitian $\eta$ leads to the conservation of probability with time, i.e.
$\frac{d}{dt}\langle\psi|{\eta}|\psi\rangle=0$, while an anti-linear anti-Hermitian $\eta$ does not in general.
Nonetheless, such a conservation law is guaranteed with respect to the anti-linear anti-Hermitian $\eta=PT$.

\item
A linear Hermitian $\eta$ leads to the unitarity of time evolution, i.e.
$\langle\psi(t)|{\eta}|\psi(t)\rangle=\langle\psi(0)|{\eta}|\psi(0)\rangle$,
where $\psi(t)=e^{-iHt}\psi(0)$,
while an anti-linear anti-Hermitian $\eta$ does not in general.
Nonetheless, such a unitarity is guaranteed with respect to the  anti-linear anti-Hermitian $\eta=PT$.

\item
A linear Hermitian $\eta$ leads to a real average value, $\overline{\langle A\rangle}_{\rm Av}=\langle A\rangle_{\rm Av}$,
for any physical observable $A$ which satisfies eq.~(\ref{pseudoH}), while an anti-linear anti-Hermitian $\eta$ does not in general.
Quite interestingly, our Hamiltonian eq.~(\ref{hamilton}) has a
real spectrum, which will be seen from the equivalence of our model and the Pais-Uhlenbeck oscillator in section 4.

\item
A linear Hermitian $\eta$ leads to the $\eta$-pseudo-Hermitian symmetry at any time: $A^{\ddag}(t)=A(t)$,
for any physical observable $A$ which satisfies eq.~(\ref{pseudoH}) at the initial time, where
$A(t)=e^{+iHt}A(0)e^{-iHt}$, while an anti-linear anti-Hermitian $\eta$ does not in general.
Nonetheless, such a symmetry is guaranteed with respect to the  anti-linear anti-Hermitian $\eta=PT$.

\item
A linear Hermitian $\eta$ leads to the usual equation of motion for the average value of $A^{\ddag}(t)$,
$\frac{d}{dt}\langle A^{\ddag}(t)\rangle_{\rm Av}=i\langle [H, A^{\ddag}(t)]\rangle_{\rm Av}$, if $\eta$ and $H$ do not explicitly contain time,
while it is meaningless to derive such an equation for an anti-linear anti-Hermitian $\eta$ because the property No.5 as a basis no longer exists.
Nonetheless, this equation is satisfied with respect to the  anti-linear anti-Hermitian $\eta=PT$.
\end{enumerate}

\section{Equivalence of our model and Pais-Uhlenbeck oscillator}

In order to establish the relationship between our mode and the Pais-Uhlenbeck oscillator, we first derive the equation of motion that the
Hamiltonian eq.~(\ref{hamilton}) corresponds to and then compare this equation of motion with that of the Pais-Uhlenbeck oscillator. To this end,
by using eq.~(\ref{hamilton}) and eq.~(\ref{canonicalRe}) we obtain the following Hamilton equations for the canonical pairs of coordinates
and momenta $x_j$ and $p_j$ from the general formulations $\dot{x}_j=i[H, {x}_j]$ and $\dot{p}_j=i[H, {p}_j]$, where $j=1,2$,
\begin{equation}
\dot{x}_1=p_1+i\frac{a_3}{2a_1a_2}p_2, \qquad \dot{x}_2=p_2+i\frac{a_3}{2a_1a_2}p_1, \nonumber
\end{equation}
\begin{equation}
\dot{p}_1=-a_1^2x_1, \qquad  \dot{p}_2=-a_2^2x_2.\label{HamiltonEOM}
\end{equation}
Eliminating the momenta $p_j$ in eq.~(\ref{HamiltonEOM}), we have the equations of motion for the coordinates $x_j$,
\begin{eqnarray}
\ddot{x}_1=-a_1^2x_1-i\frac{a_2a_3}{2a_1}x_2,\qquad \ddot{x}_2=-a_2^2x_2-i\frac{a_3a_1}{2a_2}x_1.\label{couplingEOM}
\end{eqnarray}
Again eliminating the coupling between the two spatial dimensions in eq.~(\ref{couplingEOM}),
that is, increasing the order of the derivative with respect to {\em time} up to the fourth,
we finally achieve the desired equation of motion,
\begin{equation}
\frac{d^4 x_j}{dt^4}+\left(a_1^2+a_2^2\right)\frac{d^2 x_j}{dt^2}+\left(a_1^2a_2^2+\frac{a_3^2}{4}\right)x_j=0, \qquad j=1,2.\label{PUEOM}
\end{equation}

We can see that eq.~(\ref{PUEOM}) looks very much like~\cite{PU,BM1,BM2} the equation of motion of the Pais-Uhlenbeck oscillator but contains one more
parameter $a_3$, or equivalently one more term $\frac{a_3^2}{4}x_j$.
Our model has three independent non-vanishing parameters while the Pais-Uhlenbeck oscillator has two. We shall see that our model includes
the Pais-Uhlenbeck oscillator as a special case. In order to make a precise comparison
of the two models, we introduce two new parameters $\omega_1$ and $\omega_2$ defined as follows:
\begin{eqnarray}
\omega_1^2+\omega_2^2:=a_1^2+a_2^2, \qquad \omega_1^2 \omega_2^2:=a_1^2a_2^2+\frac{a_3^2}{4}.\label{NewP}
\end{eqnarray}
It is obvious that the new parameters are just the frequencies in the Pais-Uhlenbeck oscillator if they have real solutions of the above
quadratic algebraic equation.
Solving eq.~(\ref{NewP}), we get the following three different cases.

Case I: If the three non-vanishing constants comply with the inequality,
\begin{eqnarray}
|a_3| < |a^2_1-a^2_2|,\label{condition1}
\end{eqnarray}
the two solutions are real and unequal,
\begin{eqnarray}
\omega_1^2 &=& \frac{1}{2}\left[a_1^2+a_2^2 \pm \sqrt{\left(a_1^2-a_2^2\right)^2-a_3^2}\right], \nonumber \\
\omega_2^2 &=& \frac{1}{2}\left[a_1^2+a_2^2 \mp \sqrt{\left(a_1^2-a_2^2\right)^2-a_3^2}\right].\label{Solution1}
\end{eqnarray}
For this case our model is equivalent to the Pais-Uhlenbeck oscillator with unequal frequencies~\cite{PU,BM1}.

Case II: If the three non-vanishing constants comply with the equality,
\begin{eqnarray}
|a_3| = |a^2_1-a^2_2|, \label{condition2}
\end{eqnarray}
the two solutions are real and equal,
\begin{eqnarray}
\omega_1^2 = \omega_2^2 = \frac{1}{2}\left(a_1^2+a_2^2\right).\label{Solution2}
\end{eqnarray}
For this case our model is equivalent to  the Pais-Uhlenbeck oscillator with the equal frequency~\cite{PU,BM2}.

Case III: If the three non-vanishing constants comply with the inequality,
\begin{eqnarray}
|a_3| > |a^2_1-a^2_2|, \label{condition3}
\end{eqnarray}
the two solutions are a pair of complex conjugate numbers,
\begin{eqnarray}
\omega_1^2 &=& \frac{1}{2}\left[a_1^2+a_2^2 \pm i\sqrt{a_3^2-\left(a_1^2-a_2^2\right)^2}\right], \nonumber \\
\omega_2^2 &=& \frac{1}{2}\left[a_1^2+a_2^2 \mp i\sqrt{a_3^2-\left(a_1^2-a_2^2\right)^2}\right].\label{Solution3}
\end{eqnarray}
For this case our model is beyond the region of the Pais-Uhlenbeck oscillator where $\omega_1$ and $\omega_2$ must be real and positive.
It is interesting to investigate whether our model has real spectra
in this case. However, this problem has nothing to do with the spontaneous breaking of permutation symmetry focused on in the present paper,
we thus leave it for our further consideration in a separate work.

As a whole, our model covers the Pais-Uhlenbeck oscillator. Note that the Pais-Uhlenbeck oscillator can be described by a $PT$-symmetric quantum
system~\cite{BM1,BM2}. Here we mention that the Pais-Uhlenbeck oscillator
can also be depicted by a $PT$-pseudo-Hermitian quantum system like eq.~(\ref{hamilton}).
Therefore, we provide an alternative possibility for describing the Pais-Uhlenbeck oscillator by using a different non-Hermitian Hamiltonian
from that given by refs.~\cite{BM1,BM2}.

\section{Permutation and its spontaneous breaking in our model and identity and its spontaneous breaking in the Pais-Uhlenbeck oscillator}

In this section, we first analyze in our model the permutation symmetry and its spontaneous breaking,
and then reveal how the latter has a close relationship with real energy levels free of the oscillation between positive and negative values.
Next, by using the modified imaginary-scaling scheme or its equivalent indefinite-metric scheme given in
ref.~\cite{Ali2} with which the negative norms or ghost states are circumvented for the Pais-Uhlenbeck oscillator,
we can obtain the real spectrum that is bounded below for our model in Case I. Finally,
we establish the connection of our model and the
Pais-Uhlenbeck oscillator in the aspects of the permutation and identity and of their spontaneous breaking,
which, as a byproduct, can be applied to explain the unequal-frequency condition imposed  as a prerequisite upon the Pais-Uhlenbeck oscillator in
refs~\cite{BM1,BM2,Ali2}.
Note that our model is related to
the permutation symmetry and its spontaneous breaking only in Case I.
We shall see that the permutation in our model corresponds to the identity (not in magnitude but in attribute) of $\omega_1$ and $\omega_2$
in the Pais-Uhlenbeck oscillator.
If $\omega_1=\omega_2$ is given as in Case II, the identity will become trivial and meaningless.

It is quite obvious that the Hamiltonian eq.~(\ref{hamilton}) is invariant under the following permutation transformation,
\begin{equation}
a_1\rightarrow a_2, \quad a_2\rightarrow a_1; \quad x_1\rightarrow x_2, \quad x_2\rightarrow x_1; \quad p_1\rightarrow p_2, \quad p_2\rightarrow p_1.
\label{permu}
\end{equation}
In order to investigate the relation between the permutation symmetry and a real spectrum free of interchange of positive and negative energy levels
in our model,
we diagonalize the Hamiltonian eq.~(\ref{hamilton}) by
utilizing an approach similar to that adopted in ref.~\cite{NSCC}.\footnote{We point out that the claim in ref.~\cite{NSCC}
``A non-Hermitian Hamiltonian can be changed into a
Hermitian one by a linear transformation even if related non-Hermitian terms do not vanish" is incorrect. The Hamiltonian discussed in ref.~\cite{NSCC}
looks like a Hermitian one after a linear transformation but in fact is still non-Hermitian. The non-Hermiticity has actually been hidden in
some non-Hermitian operators defined through the linear transformation. We note that the non-Hermitian Hamiltonian in ref.~\cite{NSCC} has the
same permutation symmetry as that of our Hamiltonian eq.~(\ref{hamilton}). As a consequence,
the reason that the non-Hermitian Hamiltonian in ref.~\cite{NSCC}
possesses real eigenvalues free of the oscillation between positive and negative values
is not that it can be changed into a Hermitian one but probably that the spontaneous breaking
of the symmetry happens to it. See the discussions below for the details.}
For Case I, after carefully analyzing this Hamiltonian which has the ${PT}$-pseudo-Hermiticity as eq.~(\ref{PTpseudo}),
we introduce the intermediate phase space variables, $X_1^{\prime}$, $P_1^{\prime}$, $X_2^{\prime}$, $P_2^{\prime}$,
through the linear transformations as follows:
\begin{eqnarray}
X_1^{\prime} =  \frac{a_1 x_1-\alpha_2 a_2 x_2}{\alpha_2-\alpha_1},\qquad
P_1^{\prime} = \frac{\alpha_1}{a_1}p_1+\frac{1}{a_2}p_2;\nonumber\\
X_2^{\prime} =  \frac{a_1 x_1-\alpha_1 a_2 x_2}{\alpha_1-\alpha_2}, \qquad
P_2^{\prime} = \frac{\alpha_2}{a_1}p_1+\frac{1}{a_2}p_2,\label{Newvariable}
\end{eqnarray}
and then obtain the seemingly diagonalized form of the Hamiltonian (eq.~(\ref{hamilton}))
in terms of such phase space variables as eq.~(\ref{Newvariable}),
\begin{eqnarray}
H &=& \frac{1}{2}U^{-2}\omega^2\,\frac{{P^\prime_1}^2}{1+\alpha^2_1}+\frac{1}{2}\left(1+\alpha^2_1\right){X^\prime_1}^2\nonumber\\
& & +\frac{1}{2}U^2\omega^2\,\frac{{P^\prime_2}^2}{1+\alpha^2_2}+\frac{1}{2}\left(1+\alpha^2_2\right){X^\prime_2}^2,\label{interhamilton}
\end{eqnarray}
where two pure imaginary parameters\footnote{It is inevitable to introduce the pure imaginary parameters due to
the existence of the pure imaginary interaction in the Hamiltonian eq.~(\ref{hamilton}).} $\alpha_1$ and $\alpha_2$ are defined by
\begin{eqnarray}
\alpha_1 & := & \frac{a^2_1-a^2_2+\sqrt{(a^2_1-a^2_2)^2-a^2_3}}{ia_3},\nonumber \\
\alpha_2 & := &
\frac{a^2_1-a^2_2-\sqrt{(a^2_1-a^2_2)^2-a^2_3}}{ia_3},\label{alpha}
\end{eqnarray}
which are not independent of each other but constrained by the relation:
\begin{eqnarray}
1+\alpha_1 \alpha_2=0,\label{alpha1alpha2}
\end{eqnarray}
and two real and positive parameters $U$ and $m$ are defined by
\begin{eqnarray}
U & := & \sqrt[4]{\frac{a^2_1+a^2_2-\sqrt{(a^2_1-a^2_2)^2-a^2_3}}{a^2_1+a^2_2+\sqrt{(a^2_1-a^2_2)^2-a^2_3}}},\label{U}\\
\omega& := &\sqrt[4]{a^2_1a^2_2+{a^2_3}/{4}}.\label{omega}
\end{eqnarray}

Eq.~(\ref{interhamilton}) looks like the Hamiltonian of two separated one-dimensional oscillators but in fact it is not exactly the case
because $X^\prime_j$ and $P^\prime _j$ ($j=1,2$) are not Hermitian and in particular $1+\alpha^2_1$ and $1+\alpha^2_2$
are not positive definite simultaneously.
It is easy to prove that eq.~(\ref{interhamilton}) still has the $PT$-pseudo-Hermitian symmetry as its original form eq.~(\ref{hamilton}),
that is, such a symmetry is never
altered by the linear transformation eq.~(\ref{Newvariable}). Similarly, a non-Hermitian Hamiltonian cannot be changed into a Hermitian one by a
linear transformation.
This explains how that incorrect claim occurs in ref.~\cite{NSCC},
see also footnote 6.
That $1+\alpha^2_1$ and $1+\alpha^2_2$  in eq.~(\ref{interhamilton}) cannot be positive simultaneously gives the subtlety
that the $PT$-pseudo-Hermitian quantum system possesses, which leads to the
mechanism of the spontaneous breaking of permutation symmetry. Under the permutation eq.~(\ref{permu}),
the newly introduced parameters (eq.~(\ref{alpha})) and phase space variables (eq.~(\ref{Newvariable})) transform as follows:
\begin{equation}
\alpha_1\rightarrow -\alpha_2, \quad \alpha_2\rightarrow -\alpha_1; \quad X^\prime_1\rightarrow \alpha_1X^\prime_1,
\quad X^\prime_2\rightarrow \alpha_2X^\prime_2; \quad P^\prime_1\rightarrow \frac{1}{\alpha_1}P^\prime_1,
\quad P^\prime_2\rightarrow \frac{1}{\alpha_2}P^\prime_2.\label{permu2}
\end{equation}
We can verify that eq.~(\ref{interhamilton}) is also invariant under the transformation eq.~(\ref{permu2}) which is induced by the permutation
eq.~(\ref{permu}).

Next, we analyze how the permutation symmetry is broken spontaneously.
Because $(1+\alpha^2_1)/(1+\alpha^2_2)=\alpha^2_1=1/\alpha^{2}_2 < 0$, see eq.~(\ref{condition1}), eq.~(\ref{alpha}) and eq.~(\ref{alpha1alpha2}),
the positivity of $1+\alpha^2_1$ or of $1+\alpha^2_2$
in eq.~(\ref{interhamilton}) is indefinite, but takes either of the two cases:
\begin{eqnarray}
{\rm Case \;(1):} \qquad 1+\alpha^2_1>0, \qquad 1+\alpha^2_2<0,\label{case1}
\end{eqnarray}
\begin{eqnarray}
{\rm Case \;(2):} \qquad 1+\alpha^2_1<0, \qquad 1+\alpha^2_2>0.\label{case2}
\end{eqnarray}
Although they are equivalent in dynamics, the non-diagonalized Hamiltonian eq.~(\ref{hamilton}) and its seemingly diagonalized form
eq.~(\ref{interhamilton}) have a crucial difference in formalism. This difference can be applied to the analysis of
the spontaneous breaking of permutation symmetry. That is the reason why we have to diagonalize eq.~(\ref{hamilton}). For the details, see below.
In the former (eq.~(\ref{hamilton})) the permutation symmetry has nothing to do with the given values
of the parameters $a_1$, $a_2$ and $a_3$ that satisfy the inequality eq.~(\ref{condition1}). However, in the latter (eq.~(\ref{interhamilton}))
case (1) or case (2) is definitely chosen for some given values
of the three parameters, that is, only one case is kept while the other is lost.
Although the permutation symmetry maintains for either case (1) or case (2) in eq.~(\ref{interhamilton}), it is such a difference that will lead to
the spontaneous breaking of permutation symmetry. In order to see this clearly, let us
rewrite eq.~(\ref{interhamilton}) for the two cases as follows:
\begin{eqnarray}
H^{(1), \,(2)} &=& \pm U^{-1}\left[\frac{1}{2m}\left(\sqrt{\frac{U^{-1}}{|\alpha_1|}}P^\prime_1\right)^2
+\frac{1}{2}m\omega^2\left(\sqrt{\frac{|\alpha_1|}{U^{-1}}}X^\prime_1\right)^2
\right] \nonumber\\
& & \mp U\left[\frac{1}{2m}\left(\sqrt{\frac{U}{|\alpha_2|}}P^\prime_2\right)^2
+\frac{1}{2}m\omega^2\left(\sqrt{\frac{|\alpha_2|}{U}}X^\prime_2\right)^2
\right],\label{interhamilton2}
\end{eqnarray}
where the upper and lower signs correspond to case (1) and case (2), respectively, and the real and positive parameter $m$ is defined by
\begin{eqnarray}
m := 2{\omega}^{-2}\sqrt{\left(a^2_1-a^2_2\right)^2/{a^2_3}-1}.\label{m}
\end{eqnarray}
Eq.~(\ref{interhamilton2}) is a key step for us to change eq.~(\ref{interhamilton}) into a completely diagonalized form of the Hamiltonian
eq.~(\ref{hamilton}).
We can see that the permutation symmetry is now spontaneously broken in eq.~(\ref{interhamilton2}). Under the transformation (eq.~(\ref{permu2})),
case (1) and case (2) exchange to each other, and the
Hamiltonian (eq.~(\ref{interhamilton2})) corresponding to the upper sign $H^{(1)}$ changes to that corresponding to the lower sign $H^{(2)}$,
and vice versa.
It seems to be a difficulty in the $PT$-pseudo-Hermitian quantum mechanics that the exchange of the Hamiltonians of the two cases,
see eq.~(\ref{interhamilton2}),
gives rise to the interchange between positive energy levels and negative ones. Fortunately, the spontaneous breaking of permutation symmetry
makes the quantum system escape from this obstacle, that is, the permutation symmetry is now spontaneously broken in the individual
$H^{(1)}$ or $H^{(2)}$
and no exchange between them will occur.
We may have some similarity if we compare the spontaneous breaking of permutation symmetry in our case
with the spontaneous breaking of vacuum symmetry in the Higgs mechanism of gauge field theory.
That is to say, the former chooses one branch of the $PT$-pseudo-Hermitian Hamiltonian eq.~(\ref{interhamilton2}), $H^{(1)}$ or $H^{(2)}$,
for giving a real spectrum free of the interchange of positive and negative levels
while the latter one branch of vacuum states for
producing  massive gauge bosons.

Now we introduce the final phase space variables, $X_1$, $P_1$, $X_2$, $P_2$, 
\begin{eqnarray}
X_1 = \sqrt{\frac{|\alpha_1|}{U^{-1}}}X^\prime_1, \qquad
P_1 = \sqrt{\frac{U^{-1}}{|\alpha_1|}}P^\prime_1;\nonumber\\
X_2 = \sqrt{\frac{|\alpha_2|}{U}}X^\prime_2,\qquad
P_2 = \sqrt{\frac{U}{|\alpha_2|}}P^\prime_2,
\label{Finalvariable}
\end{eqnarray}
and rewrite eq.~(\ref{interhamilton2}) in such a way that it looks like the standard formulation of a harmonic oscillator Hamiltonian,
\begin{equation}
H^{(1), \,(2)}=\pm U^{-1}\left(\frac{P^2_1}{2m}+\frac{1}{2}m\omega^2
X^2_1\right)\mp U\left(\frac{P^2_2}{2m}+\frac{1}{2}m\omega^2X^2_2\right).\label{Finalhamilton}
\end{equation}
By using eqs.~(\ref{canonicalRe}), (\ref{Newvariable}) and (\ref{Finalvariable}),
we can prove that the final phase space variables, though non-Hermitian, $X_j\neq X^{\dagger}_j$ and $P_j\neq P^{\dagger}_j$,
satisfy the same Heisenberg commutation relations as the Hermitian phase space variables $(x_j, p_j)$,
\begin{equation}
[X_j,P_k]=i{\delta}_{jk}, \qquad [X_j,X_k]=0=[P_j,P_k], \qquad
j,k=1,2.\label{XPrel}
\end{equation}
Naively, the energy spectrum of our model in Case I (see eq.~(\ref{condition1})) seems to be
\begin{equation}
E^{{\prime}\,(1), \,(2)}= \pm
\left(n_1+\frac{1}{2}\right)U^{-1}\omega \mp \left(n_2+\frac{1}{2}\right)U\omega, \qquad n_1,n_2=0,1,2,\cdots.\label{intereigenvalue}
\end{equation}
Note that for either case (1) (see eq.~(\ref{case1})) or case (2) (see eq.~(\ref{case2})) the
negative eigenvalues appear in $E^{{\prime}\,(1)}$ or $E^{{\prime}\,(2)}$.
As our model in Case I is equivalent to
the unequal-frequency Pais-Uhlenbeck oscillator (see section 3 for the detailed discussion), the appearance of negative eigenvalues represents the
appearance of negative norms or ghost states in this fourth-order derivative model.
It was regarded for a long time as a puzzling problem for the Pais-Uhlenbeck oscillator~\cite{PU},
however, this problem has been solved recently first by the imaginary-scaling scheme~\cite{BM1,BM2} and then alternatively
by the modified imaginary-scaling scheme or its equivalent indefinite-metric scheme~\cite{Ali2}.
As a result, by applying the modified imaginary-scaling scheme or its equivalent indefinite-metric scheme~\cite{Ali2} to our model
(see eq.~(\ref{hamilton}) or  eq.~(\ref{Finalhamilton})) in Case I (see eq.~(\ref{condition1})) for the both case (1) and case (2) we obtain
the desired spectrum which is free of negative values or which is bounded below,
\begin{equation}
E= \left(n_1+\frac{1}{2}\right)U^{-1}\omega + \left(n_2+\frac{1}{2}\right)U\omega, \qquad n_1,n_2=0,1,2,\cdots.\label{eigenvalue}
\end{equation}
We mention that the spectrum is independent of whether case (1) or case (2) is taken. That is, the positivity of eigenvalues is
independent of the values taken for the three non-vanishing parameters $a_1$, $a_2$ and $a_3$ constrained by the inequality eq.~(\ref{condition1}).
This outcome is reasonable because the original formulation of our Hamiltonian eq.~(\ref{hamilton}) under the constraint eq.~(\ref{condition1})
should have a unique spectrum. As to the spontaneous breaking of permutation symmetry which has played a crucial role in giving a real spectrum free of
interchange between positive and negative eigenvalues, it is just a tool or a mechanism with which we can achieve our goal.

Now we turn to the connection of our model and the Pais-Uhlenbeck oscillator. According to eqs.~(\ref{PUEOM})-(\ref{Solution1}),
the equation of motion of our model coincides exactly with the standard formulation of
the equation of motion of the Pais-Uhlenbeck oscillator,
\begin{equation}
\frac{d^4 x_j}{dt^4}+\left(\omega_1^2+\omega_2^2\right)\frac{d^2 x_j}{dt^2}+\omega_1^2\omega_2^2x_j=0, \qquad j=1,2.\label{PUEOM2}
\end{equation}
For the unequal-frequency case,
$\omega_1 \neq \omega_2$, this equation of motion is invariant under the
permutation:\footnote{For the equal-frequency case, such an invariance is trivial.}
$\omega_1 \rightarrow \omega_2$, $\omega_2 \rightarrow \omega_1$. In particular,
we point out that one cannot distinguish one of the two frequencies
is larger or smaller than the other. Here this property is called {\em identity}
of the two frequencies in attribute.
Eq.~(\ref{Solution1}) explicitly presents such an identity, i.e., for the upper sign $\omega_1 > \omega_2$, while for the
lower sign $\omega_1 < \omega_2$.
However, as stated in refs.~\cite{BM1,BM2,Ali2}, one of the signs must be chosen as a prerequisite
in order to calculate the spectrum of the unequal-frequency
Pais-Uhlenbeck oscillator, i.e., the unequal-frequency condition, for instance, $\omega_1 > \omega_2$, should be imposed.
In other words, the identity of the unequal $\omega_1$ and $\omega_2$ must be broken spontaneously, i.e., one has to distinguish one frequency
is larger or smaller than the other. For the choice $\omega_1 > \omega_2$,  we get the relation of the parameters $U$ and $\omega$ in our model and
the frequencies $\omega_1$ and $\omega_2$ in the Pais-Uhlenbeck oscillator by using eqs.~(\ref{Solution1}), (\ref{U}) and (\ref{omega}),
\begin{eqnarray}
& & \omega_1=U^{-1}\omega =  \sqrt{\frac{a_1^2+a_2^2 + \sqrt{\left(a_1^2-a_2^2\right)^2-a_3^2}}{2}}, \\
& & \omega_2=U\omega = \sqrt{\frac{a_1^2+a_2^2 - \sqrt{\left(a_1^2-a_2^2\right)^2-a_3^2}}{2}}.\label{conn}
\end{eqnarray}
Therefore, the spectrum of our model eq.~(\ref{eigenvalue}) can be rewritten as
\begin{equation}
E= \left(n_1+\frac{1}{2}\right)\omega_1 + \left(n_2+\frac{1}{2}\right)\omega_2, \qquad n_1,n_2=0,1,2,\cdots,\label{eigenvalue2}
\end{equation}
which is just the formulation given in refs.~\cite{BM1,BM2,Ali2}.

As a summary in this section, we emphasize the connection of our model and the Pais-Uhlenbeck oscillator in the following two aspects. First,
the former has the permutation (symmetry) of two dimensions, while the latter the identity (a class of symmetry) of the unequal
frequencies $\omega_1$ and $\omega_2$. Second, the permutation in the former is broken spontaneously, while the identity in the latter
is broken spontaneously. The spontaneous breaking of identity gives a reasonable explanation for the unequal-frequency condition
imposed as a prerequisite upon the oscillator model
in refs.~\cite{BM1,BM2,Ali2}. Therefore, we may conclude that the spontaneous breaking of
symmetries\footnote{The symmetry is permutation in our model and identity in
the Pais-Uhlenbeck oscillator.} is a general phenomenon which exists in a wider region than that we already knew.

\section{Conclusion}
In this paper, due to the motivation to search for a pseudo-Hermitian quantum system that can describe the Pais-Uhlenbeck oscillator,
we construct a concrete model (eq.~(\ref{hamilton})) which has the $PT$-pseudo-Hermitian symmetry.
As our model belongs to the region of
the $\eta$-pseudo-Hermitian quantum mechanics associated with an anti-linear anti-Hermitian $\eta$,
we summarize the characteristics that such a system normally possesses,
see section 2 and the Appendix for the details, where they give us particular interest when $\eta$ takes the specific $PT$. By deriving the
equation of motion for our Hamiltonian, we explicitly show that our model covers the Pais-Uhlenbeck oscillator as a special case. In particular,
we point out that our model is invariant under the permutation of two dimensions and that such an invariance (symmetry) should be broken
spontaneously in order to obtain a real spectrum that is free of interchange between positive and negative eigenvalues. Moreover, by comparing
our model with the Pais-Uhlenbeck oscillator, we reveal that the permutation of two dimensions in our model corresponds to
the identity of two frequencies in the unequal-frequency Pais-Uhlenbeck oscillator, and therefore that the spontaneous breaking of
permutation in our model corresponds to the spontaneous breaking of identity in the unequal-frequency Pais-Uhlenbeck oscillator. We want to
emphasize that the choice $\omega_1 > \omega_2$ or $\omega_1 < \omega_2$, though without loss of generality, was made as a prerequisite in
refs.~\cite{BM1,BM2,Ali2}, here it is explained reasonably
from the point of view of
the spontaneous breaking of identity. That is, it is this mechanism of
the spontaneous breaking of identity that makes such a choice achieved. We also point out that
we can obtain the same real spectrum as that given in refs.~\cite{BM1,BM2,Ali2} for our model
with the help of the modified imaginary-scaling scheme or its equivalent indefinite-metric scheme~\cite{Ali2}.
Finally, we mention that for our model in Case II (eq.~(\ref{condition2})), i.e., the equal-frequency
Pais-Uhlenbeck oscillator has been studied~\cite{BM2} in detail
and therefore we no longer revisit it here mainly due to its independence of the spontaneous breaking of symmetry,
and that for our model in Case III (eq.~(\ref{condition3})), as analyzed under eq.~(\ref{Solution3}), it
is beyond the region of the Pais-Uhlenbeck oscillator
and also independent of the spontaneous breaking of symmetry,
we thus leave it for our further consideration in a separate work.

\section*{Acknowledgments}
Y-GM would like to thank
H.P. Nilles of the University of Bonn for kind hospitality where part of the work was performed.
This work was supported in part by the Alexander von Humboldt Foundation under a short term programme,
by the National Natural
Science Foundation of China under grant No.11175090, and by the Fundamental Research Funds for the Central Universities
under grant No.65030021.
At last, the authors would like to thank the anonymous referee for the helpful comments which indeed improve this paper greatly.

\newpage
\section*{Appendix}
We adopt the notations used in ref.~\cite{Pauli} except for the symbol of the $\eta$-pseudo-Hermitian adjoint.

\subsection*{A1: Reality of probability}
The probability, i.e. the bilinear form with respect to the indefinite metric $\eta$ defined by Pauli~\cite{Pauli} is
\begin{equation}
\int \overline{\psi}\eta{\psi} dq=\sum {\overline{\psi}}_n\eta_{nm}{\psi}_m,
\end{equation}
and its complex conjugate then takes the form,
\begin{equation}
\int \overline{(\overline{\psi}\eta{\psi})} dq=\sum \overline{({\overline{\psi}}_n\eta_{nm}{\psi}_m)}
=\sum {\psi}_n\overline{\eta}_{nm}\overline{\psi}_m.
\end{equation}
When $\eta$ is linear Hermitian, i.e. $\overline{\eta}_{nm}={\eta}_{mn}$, ${\psi}_n{\eta}_{mn}={\eta}_{mn}{\psi}_n$, and
${\eta}_{mn}\overline{\psi}_m=\overline{\psi}_m{\eta}_{mn}$, we obtain
\begin{equation}
\int \overline{(\overline{\psi}\eta{\psi})} dq = \sum {\psi}_n{\eta}_{mn}\overline{\psi}_m
=\sum \overline{\psi}_m{\eta}_{mn}{\psi}_n=\int \overline{\psi}\eta{\psi} dq.
\end{equation}
That is, the probability is definitely real. However, if $\eta$ is anti-linear anti-Hermitian, i.e.
$\eta(a|\psi_1\rangle+b|\psi_2\rangle)=\overline{a}\eta|\psi_1\rangle+\overline{b}\eta|\psi_2\rangle$ and
$\int \overline{\psi}\eta{\phi} dq=\int \overline{\phi}\eta{\psi} dq$, see footnote 3,  we have
\begin{equation}
\int \overline{(\overline{\psi}\eta{\psi})} dq
=\sum {\psi}_n \overline{\eta}_{nm}\overline{\psi}_m \neq \int \overline{\psi}\eta{\psi} dq.
\end{equation}
That is, the probability is not real in general.

Because our model with the constraint eq.~(\ref{condition1}) is equivalent to the unequal-frequency Pais-Uhlenbeck oscillator
that has been shown~\cite{BM1,BM2,Ali2} to be stable and unitary, thus
the probability with respect to the indefinite metric $PT$ must be real and positive definite.

\subsection*{A2: Conservation of probability}
From the Schr\"odinger equation, $\frac{\partial \psi}{\partial t}=-iH\psi$, we have
\begin{equation}
\overline{\psi}\eta\frac{\partial \psi}{\partial t}=-\overline{\psi}\eta i H\psi,\label{a21}
\end{equation}
and from the complex conjugate form of the Schr\"odinger equation, $\frac{\partial \overline{\psi}}{\partial t}=i\overline{\psi}H^{\dag}$, we get
\begin{equation}
\frac{\partial \overline{\psi}}{\partial t}\eta\psi=i\overline{\psi}H^{\dag}\eta\psi=i\overline{\psi}\eta H^{\ddag}\psi.\label{a22}
\end{equation}
If $\eta$ is anti-linear anti-Hermitian, eq.~(\ref{a21}) becomes
\begin{equation}
\overline{\psi}\eta\frac{\partial \psi}{\partial t}=+i\overline{\psi}\eta  H\psi.\label{a23}
\end{equation}
Note that the sign is changed on the right-hand side! Subtracting eq.~(\ref{a23}) from eq.~(\ref{a22})
and integrating the difference on the both sides in the coordinate space $q$, we obtain
\begin{equation}
\int \left(\frac{\partial \overline{\psi}}{\partial t}\eta\psi-\overline{\psi}\eta\frac{\partial \psi}{\partial t}\right) dq
=\int i\overline{\psi}\eta (H^{\ddag}-H)\psi dq.\label{a24}
\end{equation}
The right-hand side of the above equation equals zero because of $H^{\ddag}=H$. However, if $\eta$ is an anti-linear anti-Hermitian operator
that does not contain the time reversal operator $T$, the left-hand side cannot be written as $\overline{\psi}\eta{\psi}$.
This means that the probability
is not conserved in general. Fortunately, in our model $\eta=PT$ and we can reduce the left-hand side of eq.~(\ref{a24}) to be
\begin{equation}
\int \left(\frac{\partial \overline{\psi}}{\partial t}\eta\psi-\overline{\psi}\eta\frac{\partial \psi}{\partial t}\right) dq
=\int \frac{\partial}{\partial t} (\overline{\psi}\eta{\psi}) dq = \frac{d}{d t}\int \overline{\psi}\eta{\psi} dq.\label{a25}
\end{equation}
Note that because $\eta$ contains the time reversal operator the sign is changed when $\eta$ and $\frac{\partial}{\partial t}$ exchange
in the second term of the left-hand side of eq.~(\ref{a25})!  Therefore, the conservation of probability with time is ensured
in our model (eq.~(\ref{hamilton})).

\subsection*{A3: Unitarity of time evolution}
If the wavefunction at the initial time is represented by $\psi(0)$,  it takes the following form at any time according to the Schr\"odinger equation,
\begin{equation}
\psi(t)=e^{-iHt}\psi(0).
\end{equation}
We now calculate the inner product of $\psi(t)$ with respect to the indefinite metric $\eta$,
\begin{equation}
\langle \psi(t)|\psi(t)\rangle_{\eta}=\langle \psi(t)|{\eta}|\psi(t)\rangle
=\langle \psi(0)|{\eta}({\eta}^{-1}e^{+iH^{\dag}t}{\eta})e^{-iHt}|\psi(0)\rangle.\label{a41}
\end{equation}
If $\eta$ is linear Hermitian, we immediately get ${\eta}^{-1}e^{+iH^{\dag}t}{\eta}=e^{+iHt}$ by using eq.~(\ref{pseudoH}). Therefore,
eq.~(\ref{a41}) gives the unitarity of time evolution:
$\langle \psi(t)|\psi(t)\rangle_{\eta}=\langle \psi(0)|{\eta}|\psi(0)\rangle=\langle \psi(0)|\psi(0)\rangle_{\eta}$. However, if
$\eta$ is an anti-linear anti-Hermitian operator that does not contain the time reversal operation, we have
${\eta}^{-1}(+iH^{\dag}t){\eta}=-iHt$ after considering eq.~(\ref{pseudoH}), and thus we know ${\eta}^{-1}e^{+iH^{\dag}t}{\eta}\neq e^{+iHt}$, i.e.
eq.~(\ref{a41}) does not lead to the unitarity of time evolution. Quite interestingly,
if $\eta$ is an anti-linear anti-Hermitian operator that contains the time reversal operation, such as our choice $\eta=PT$,
we obtain ${(PT)}^{-1}(+iH^{\dag}t){(PT)}=+iHt$, where we have used the property ${(PT)}^{-1}(+iH^{\dag}t){(PT)}={(PT)}^{-1}H^{\dag}{(PT)}(+it)$ and
eq.~(\ref{pseudoH}).
Thus, we again obtain ${\eta}^{-1}e^{+iH^{\dag}t}{\eta}=e^{+iHt}$. That is, the unitarity of time evolution is ensured.

\subsection*{A4: Average value}
From the definition of the $\eta$-pseudo-Hermitian adjoint of an arbitrary physical observable $A$, $A^{\ddag} := \eta^{-1} A^\dagger\eta$,
the average value of $A^{\ddag}$ takes the form,
\begin{equation}
\langle A^{\ddag}\rangle_{\rm Av}=\langle \eta^{-1} A^\dagger\eta\rangle_{\rm Av}=\sum {\overline{\psi}}_n\eta_{nm}
(\eta^{-1} A^\dagger\eta)_{ml}{\psi}_l=\sum {\overline{\psi}}_n{\overline{A}}_{mn}\eta_{ml}{\psi}_l.\label{a11}
\end{equation}
If $\eta$ is linear Hermitian, the above equation becomes
\begin{equation}
\langle A^{\ddag}\rangle_{\rm Av}=\sum \overline{({\overline{\psi}}_n\eta_{nm}A_{ml}{\psi}_l)}
=\overline{\langle A\rangle}_{\rm Av}.
\end{equation}
Therefore,  it is obvious that the average value is real,
$\langle A^{\ddag}\rangle_{\rm Av}=\langle A\rangle_{\rm Av}=\overline{\langle A\rangle}_{\rm Av}$,
for any physical observable $A=A^{\ddag}$. However, if $\eta$ is anti-linear anti-Hermitian,
$\langle \psi|\eta|\phi\rangle=\langle \phi|\eta|\psi\rangle$, eq.~(\ref{a11}) reduces to be
\begin{equation}
\langle A^{\ddag}\rangle_{\rm Av}=\sum \overline{(A\psi)}_n\eta_{nm}{\psi}_m=\sum {\overline{\psi}}_n\eta_{nm}(A{\psi})_m
=\langle A\rangle_{\rm Av}.
\end{equation}
We can see that the $\eta$-pseudo-Hermitian self-adjoint condition of $A$, $A=A^{\ddag}$, does not give any restrictions to the above average value.
As a consequence, $\langle A\rangle_{\rm Av}$ is complex in general for an anti-linear anti-Hermitian $\eta$.

Due to the equivalence of our $PT$-pseudo-Hermitian model and the unequal-frequency
Pais-Uhlenbeck oscillator, our Hamiltonian has a real spectrum that is bounded below (see section 4).
This shows that the $PT$-pseudo-Hermiticity is of particular significance among general (anti-linear anti-Hermitian) $\eta$-pseudo-Hermitian symmetries.

\subsection*{A5: $\eta$-pseudo-Hermitian symmetry of observables at any time}
If an observable $A$ at the initial time is represented by $A(0)$,  it has the following form at any time according to the Heisenberg equation,
\begin{equation}
A(t)=e^{iHt}A(0)e^{-iHt}.\label{a51}
\end{equation}
We then calculate its $\eta$-pseudo-Hermitian adjoint by using eq.~(\ref{pseudoH}) and eq.~(\ref{a51}),
\begin{eqnarray}
A^{\ddag}(t)&=&{\eta}^{-1}A^{\dag}(t){\eta}={\eta}^{-1}(e^{iHt}A(0)e^{-iHt})^{\dag}{\eta}
={\eta}^{-1}e^{iH^{\dag}t}A^{\dag}(0)e^{-iH^{\dag}t}{\eta} \nonumber \\
&=&({\eta}^{-1}e^{iH^{\dag}t}{\eta})A^{\ddag}(0)({\eta}^{-1}e^{-iH^{\dag}t}{\eta}).\label{a52}
\end{eqnarray}
As discussed in A3, if $\eta$ is linear Hermitian, or if it is an anti-linear anti-Hermitian operator that contains the time reversal
operator, such as $\eta=PT$, we have ${\eta}^{-1}e^{iH^{\dag}t}{\eta}=e^{iHt}$ and ${\eta}^{-1}e^{-iH^{\dag}t}{\eta}=e^{-iHt}$. Therefore,
eq.~(\ref{a52}) reduces to be
\begin{equation}
A^{\ddag}(t)=e^{iHt}A^{\ddag}(0)e^{-iHt}.\label{a53}
\end{equation}
Considering the self-adjoint of the physical observable at the initial time, $A^{\ddag}(0)=A(0)$,
we thus prove the $\eta$-pseudo-Hermitian symmetry of $A$ at any time,
\begin{equation}
A^{\ddag}(t)=e^{iHt}A^{\ddag}(0)e^{-iHt}=e^{iHt}A(0)e^{-iHt}=A(t).
\end{equation}
Note that the above symmetry is broken in general if $\eta$ is an anti-linear anti-Hermitian operator that does not include the time reversal.

\subsection*{A6: Equation of motion for the average value of $A^{\ddag}(t)$}
From the definition of average values and eq.~(\ref{a53}), we have
\begin{equation}
\langle A^{\ddag}(t)\rangle_{\rm Av}=\langle \psi|\eta \,e^{iHt}A^{\ddag}(0)e^{-iHt}\,|\psi\rangle. \label{a61}
\end{equation}
When $\eta$ is linear Hermitian, we can derive from eq.~(\ref{a61}) the following formula,
\begin{eqnarray}
\frac{d}{d t}\langle A^{\ddag}(t)\rangle_{\rm Av} &=& \langle \psi|\eta \,iHe^{iHt}A^{\ddag}(0)e^{-iHt}\,|\psi\rangle
+\langle \psi|\eta \,e^{iHt}A^{\ddag}(0)(-iH)e^{-iHt}\,|\psi\rangle \nonumber \\
&=& i\langle [H, A^{\ddag}(t)]\rangle_{\rm Av},
\end{eqnarray}
where we have postulated that $\eta$ and $H$ do not explicitly contain time. If $\eta$ is an anti-linear anti-Hermitian operator that contains the time
reversal operator, such as $\eta=PT$, we have
\begin{eqnarray}
\frac{d}{d t}\langle A^{\ddag}(t)\rangle_{\rm Av} &=& \langle \psi|\eta \,\frac{\partial}{\partial(-t)}e^{iHt}\cdot A^{\ddag}(0)e^{-iHt}\,|\psi\rangle
+\langle \psi|\eta \,e^{iHt}A^{\ddag}(0)\frac{\partial}{\partial(-t)}e^{-iHt}\,|\psi\rangle \nonumber \\
&=& \langle \psi|\eta \,(-iH)e^{iHt} A^{\ddag}(0)e^{-iHt}\,|\psi\rangle
+\langle \psi|\eta \,e^{iHt}A^{\ddag}(0)(iH)e^{-iHt}\,|\psi\rangle \nonumber \\
&=& i\langle \psi|\eta \,HA^{\ddag}(t)\,|\psi\rangle
-i\langle \psi|\eta \,A^{\ddag}(t)H\,|\psi\rangle \nonumber \\
&=& i\langle [H, A^{\ddag}(t)]\rangle_{\rm Av}.
\end{eqnarray}
Note that we have used the properties of the anti-linear operator: $\frac{\partial}{\partial t} \eta =\eta\frac{\partial}{\partial(-t)}$
and $\eta i=-i \eta$. According to A5, the observable $A(t)$ is not $\eta$-pseudo-Hermitian self-adjoint, i.e. $A^{\ddag}(t)\neq A(t)$,
when $\eta$ is an anti-linear anti-Hermitian operator that does not contain the time
reversal operator. Therefore, we ignore this meaningless case here.

The above analysis shows that the Heisenberg equation possesses the $\eta$-pseudo-Hermitian symmetry if and only if $\eta$ is linear Hermitian
or is an anti-linear anti-Hermitian operator that contains the time reversal.

\end{document}